\documentstyle[12pt,a4]{article}

\begin{document}
\bibliographystyle{unsrt}
\input{epsf}
\topmargin 0pt
\headheight 0pt
\headsep 0pt
\textheight 9in
\topmargin 0in
 
%**********************************************************************
%GREEK LETTERS
\def\a{\alpha}
\def\b{\beta}
\def\c{\gamma}
\def\d{\delta}
\def\e{\epsilon}
\def\h{\eta}
\def\k{\kappa}
\def\l{\lambda}
\def\m{\mu}
\def\n{\nu}
\def\o{\theta}
\def\p{\pi}
\def\r{\rho}
\def\s{\sigma}
\def\t{\tau}
\def\u{\upsilon}
\def\w{\omega}
\def\x{\chi}
\def\y{xi}
\def\z{\zeta}
 
\def\C{\Gamma}
\def\D{\Delta}
\def\L{\Lambda}
\def\O{\Theta}
\def\P{\Pi}
\def\S{\Sigma}
\def\W{\Omega}
%**********************************************************************
%OTHER MACROS
 
\def\ux{\underline x}
\def\uy{\underline y}
\def\uz{\underline z}
\def\uu{\underline u}
\def\uv{\underline v}
\def\uk{\underline k}
\def\up{\underline p}
\def\pl{\partial}
\def\vphi{\varphi}
\def\rt{\rightarrow}
\def\L{\(L\) }
\def\R{\(R\) }
\def\F{\(F\) }
\def\P{\(P\) }

\newcommand{\schr} {Schr\"{o}dinger}
\newcommand{\schw} {Schwarzschild}
\def\be{\begin{equation}}
\def\ee{\end{equation}}
\def\bl{(\lambda)}
\def\ll{\lambda}
\def\tphistar{\widetilde{\phi}^\ast}
\def\tpsistar{\widetilde{\psi}^\ast}
\def\tphi{\widetilde{\phi}}
\def\tpsi{\widetilde{\psi}}
\def\phistar{\phi^\ast}
\def\psistar{\psi^\ast}
\def\tDelta{\widetilde{\Delta}}
\def\half{\frac{1}{2}} 
\def\tG{\widetilde{\cal G}} 
\newcommand{\sfrac}[2]{\mbox{$\frac{#1}{#2}$}}
%**********************************************************************
\begin{titlepage}
  {\hfill SWAT 96/124}

  {\hfill gr-qc/9607032}

  {\hfill 14 July, 1996.}
\vspace*{2cm}

\centerline{\Large{\bf The Schr\"odinger Wave Functional}}
\vspace{0.7cm}
\centerline{\Large{\bf and Vacuum States in Curved Spacetime II}}
\vspace{0.7cm}
\centerline{\Large{\bf   -- Boundaries and Foliations}}
\vspace{1.5cm}
\centerline{\bf D.V.~Long\footnote{\tt D.V.Long@swansea.ac.uk} 
\ \  and \ \ G.M.~Shore\footnote{\tt G.M.Shore@swansea.ac.uk}}
\vspace{0.8cm}
\centerline{\it Department of Physics}
\centerline{\it University of Wales Swansea}
\centerline{\it Singleton Park}
\centerline{\it Swansea, SA2 8PP, U.K. }
\vspace{1.5cm}

\begin{abstract} 
In a recent paper, general solutions for the vacuum wave functionals
in the {\schr} picture were given for a variety of classes of curved
spacetimes. Here, we describe a number of simple examples which illustrate
how the presence of spacetime boundaries influences the 
vacuum wave functional and how physical quantities are independent
of the choice of spacetime foliation used in the {\schr} approach despite
the foliation dependence of the wave functionals themselves.
\end{abstract}

\hfill
\end{titlepage}
%\tableofcontents
\baselineskip=16pt plus 0.2pt

\section{Introduction}

The {\schr} wave functional provides a simple and intuitive description
of vacuum states in quantum field theory in curved spacetimes. It is 
particularly useful in situations where the background metric
is time-dependent or in the presence of boundaries.

This is the second paper in a series where we develop the {\schr} picture
formalism in curved spacetime. In the first paper \cite{long96}, we 
reviewed and 
developed techniques for solving the {\schr} wave functional equation
for broad classes of spacetimes, viz.~static (where the metric
depends only on the spacelike coordinates), dynamic or Bianchi type I
(where the metric depends only on the timelike coordinates) and a certain
class of conformally static metrics including the Robertson-Walker 
spacetimes.  Here, we continue this development by studying examples
of spacetimes with boundaries, in particular regions described by
coordinate patches which can be analytically extended to a larger spacetime.
We describe how the presence of boundaries influences the choice of foliation
in the {\schr} formulation and determines the nature of possible
vacuum states.

The main advantage of the {\schr} picture over other ways to 
characterise vacuum states is that it describes states  
explicitly by a simple wave functional specified by a single, possibly 
time-dependent, kernel function satisfying a differential equation with the 
prescribed boundary conditions. This makes no reference to the assumed 
spectrum of excited states and so circumvents the difficulties of the 
conventional canonical description of a vacuum as a `no-particle' state 
with respect to the creation and annihilation operators
defined by a particular mode decomposition of the field, an approach 
which is not well suited to time-dependent problems. Unlike the
alternative of specifying a vacuum state implicitly by giving a prescription
for determining the Green functions, the {\schr} wave functional is an
explicit description, and this simplifies the interpretation of the nature of 
the states. In the end, of course, the same fundamental ambiguities appear
in very similar guises in all these formalisms, but while the Green function
approach is perhaps better suited to more elaborate issues such as
renormalisation and higher-order perturbative calculations, the {\schr}
picture frequently gives the clearest insight into the nature of the vacuum
state.

So far, we have spoken loosely about `the vacuum state.' In fact,
it is only for the very special class of static spacetimes that an essentially
unique state exists which possesses most of the defining attributes 
of the Minkowski vacuum. In the general case, there may be no
distinguished candidate at all for a vacuum state with the usual properties.
For example, in a dynamic spacetime, there is a one-parameter family 
of `vacuum' solutions to the {\schr} wave functional equation and the 
selection of one of these requires a physically motivated initial condition
on the first-order time-dependent equation for the kernel. Although these 
states are stable, they are not stationary states with respect to the chosen time 
evolution.

Even in a static spacetime, the vacuum wave functional will depend on the
foliation of spacetime chosen to define the {\schr} equation. On the other
hand, we expect physical observables to be independent of the choice
of foliation, given the same spacetime and boundary conditions.
The resolution of this potential paradox is illustrated here for a 
simple but non-trivial example.  

Quantum field theories in spacetimes with boundaries have been extensively
studied elsewhere \cite{fulling73, candelas76, lee86}.
In particular, questions of renormalisation and the
{\schr} picture have been addressed in considerable generality in 
\cite{mcavity93}. In this paper, our approach is rather to illustrate general
features in a number of simple and clear examples.

The content of this paper is as follows. In section 2, we review very briefly
the solutions of the {\schr} wave functional equation found in \cite{long96}.
In section 3, we continue the development of \cite{long96} by looking
at vacuum solutions in the Milne universe, an example of a dynamic spacetime 
of Robertson-Walker type which expands from an initial point
but has no asymptotically
static region. This is also of interest as an example of a spacetime
which is just one coordinate patch of a larger spacetime, the covering
spacetime in this case being simply Minkowski.
It has also found a recent application in the dynamics of bubble nucleation
in certain variants of the inflationary universe scenario \cite{hamazaki96}.

In section 4, we consider the much-studied Rindler wedge, imposing
vanishing boundary conditions on the field. The interpretation of the 
vacuum state defined with respect to a foliation respecting these
boundary conditions is considered in some detail. 

Taken together, two Rindler wedges and the Milne universe and its 
time-reversed counterpart comprise standard Minkowski spacetime.
In section 5, we describe conventional Minkowski field theory using
the foliation appropriate to the Rindler-Milne tiling and verify that,
given the correct implementation of boundary conditions, the conventional
Minkowski Green functions are recovered. This is strong evidence for the
expected foliation independence of physical observables and an
important consistency check on our interpretation of the {\schr} picture 
formalism.

This example also serves as a technical warm-up for our eventual goal
of determining the vacuum wave functional in the Kruskal black hole
spacetime, which shares many features of the Rindler-Milne foliation
of Minkowski spacetime (see, e.g. \cite{boulware75}).

\section{Vacuum Wave Functionals}

We begin by reviewing briefly the vacuum wave functional solutions 
described in \cite{long96} for different classes of spacetime.
For notation and conventions, see ref.\cite{long96}.

We consider globally hyperbolic spacetimes ${\cal M}$, with metric
$g_{\m\n}$, which admit a foliation into a family of spacelike
hypersurfaces ${\S}$, with intrinsic coordinates $\xi^i$, labelled by a 
`time' parameter $s$.
The embeddings of $\S$ in ${\cal M}$ are specified by the spacetime
coordinates $x^\m(s,\xi^i)$. 

States are described by  wave functionals $\Psi[\phi({\underline \xi}),s;
N, N^i, h_{ij}]$, where the variables $\phi({\underline \xi})$ are 
eigenvalues of the field operator on the equal-$s$ hypersurfaces $\S$,
and the {\schr} equation describes their evolution along the integral
curves of $s$.
$N$, $N^i$ and $h_{ij}$ are respectively the lapse and shift functions
characterising the embedding and the induced metric on $\S$,
The {\schr} equation for a free massive scalar field theory is then
\begin{equation}  
i{\frac {\pl\Psi}{\pl s}}= ~ \int_{\S}
{d^d\underline{\xi}}~\Bigl\{ \half N \sqrt{-h} \Bigl( \frac{1}{h}
\frac{\delta ^2}{\delta \phi ^2} -h^{ij}\pl_{i}
\phi\pl_j\phi+(m^2+\xi R)\phi^2 \Bigr) - i N^i \pl_i\phi
\frac{\delta}{\delta \phi}\Bigr\} \Psi
\label{eq:swfemanifest}
\end{equation}

While this equation makes the dependence of the wave functional on the 
foliation explicit, it is much simpler in particular examples to choose
spacetime coordinates which reflect the foliation. If we identify the 
spacetime coordinates $(t, \ux)$ with the embedding variables 
$(s,\underline \xi)$, the lapse and shift functions reduce to $N= \sqrt{g_{00}}$
and $N^i = 0$ (so that $g_{0i} = 0$) while $h_{ij} = g_{ij}$.
The {\schr} equation then reduces to 
\be
 i{\frac {\pl\Psi}{\pl t}}~=~  \half \int{d^d\ux}~
  \sqrt{-g} ~ \Bigl\{ {g_{00}\over g}
  \frac{\delta ^2}{\delta \phi ^2}  -   g^{ij}(\pl_{i}
  \phi)(\pl_{j}\phi)+(m^2+ \xi R)\phi^2  \Bigr\} \Psi
  \label{eq:swfecst}
\ee

The `vacuum' solutions to the {\schr} equation are Gaussian functionals
\be
  \Psi_0[\phi(\ux),t]=N_0(t)\psi_0[\phi,t]
  \label{eq:wavecst}
\ee
with
\be
 \psi_0[\phi,t]=  \exp\left\{{-{1\over2}   \int d^d\ux
    \sqrt{-h_x}  \int d^d\uy    \sqrt{-h_y} ~
    \phi(\ux)   G(\ux, \uy ;t) \phi(\uy)}\right\}
\label{eq:wave2cst}
\ee
and
\be
  \frac{d \ln N_0(t)}{dt} = - {i\over2} \int d^d\ux ~
  \sqrt{-h_x} \sqrt{g_{00}^x} ~ G(\ux,\ux;t)
\label{eq:timecst}
\ee
where the kernel $G(\ux,\uy;t)$ satisfies\footnote{The $(d+1)$ dimensional 
spacetime Laplacian $\Box = \frac{1}{\sqrt{-g}}\partial_\mu
(g^{\mu \nu}\sqrt{-g} \partial_\nu)$ can be split into a spatial part,
$\Box_i$ and a time part, $\Box_0$. The delta function density is given by
\( \d^d(\ux,\uy) = (\sqrt{-h_x})^{-1} \d^d(\ux - \uy) \)
and satisfies \(\int d^d\ux \sqrt{-h_x} \d^d(\ux,\uy)
f(\ux) = f(\uy)\).}
\begin{eqnarray}
  i\frac{\pl}{\pl t}\left( \sqrt{h_x h_y} ~
    G(\ux,\uy;t) \right) &=& ~
  \int{d^d\uz ~ \sqrt{-h_z}\sqrt{g_{00}^z}
   ~\sqrt{h_x h_y}}
  ~ G(\ux,\uz;t)G(\uz,\uy;t)
  \nonumber \\ && \nonumber \\ &&- ~
  \sqrt{h_x h_y} \sqrt{g_{00}^x}
   ~(\Box_i + m^2 + \xi R)_x ~
  \delta^d(\ux ,\uy)
\label{eq:kernelcst}
\end{eqnarray}

The kernel equation can be solved explicitly for special classes of spacetime.
For `static' spacetimes, where the metric depends only on the spacelike
coordinates, the kernel (which in this case is time independent) is
\be
  G(\ux,\uy)=\sqrt{g^{00}_x\,g^{00}_y} \int \frac{d\m(\l)}{(2\pi)^d} 
  ~\w(\l)~  \tpsi_{(\l)}(\w,\ux)\tpsistar_{(\l)}(\w,\uy)
\label{eq:kerstatic}
\ee
where $\tilde \psi_{(\l)}(\w, \ux)$ are a complete, orthonormal set
of solutions to the eigenvalue equation
\be
  (\Box_i +  m^2 +\xi R)\tpsi_{(\l)}(\w,\ux)=
g^{00}\w^2(\l) \tpsi_{(\l)}(\w,\ux)
\label{eq:eigenstatic}
\ee
and $d\m(\l)$ is the appropriate measure.

For `dynamic' (Bianchi type I) spacetimes, where the metric depends 
only on the time coordinate, the kernel is
\be
G(\ux,\uy;t) = -i \sqrt{g^{00}}
{1\over \sqrt{-h}}
\int {d^d\uk\over (2\pi)^d} e^{i \uk.(\ux-\uy)}~
{\pl\over\pl t} {\rm ln} \tpsi^\ast(t,\uk)
\label{eq:kerndynamic}
\ee
where $\tilde \psi(t,\uk)$ satisfies the Fourier-transformed wave equation
\be
(\Box_0 - g^{ij}k_ik_j + m^2 + \xi R) \tpsi(t,\uk) = 0
\label{eq:box0}
\ee
The arbitrariness in the choice of solution is responsible for the 
one-parameter ambiguity (strictly, a one-function ambiguity, since
the arbitrary coefficients in the general solution of
eq.(\ref{eq:box0}) may be functions of the momentum $k$)
 of the vacuum wave functional for 
dynamic spacetimes. It is important to notice that despite the time-dependence
of the kernel, the vacuum states described by eq.(\ref{eq:kerndynamic}) are 
stable
and can allow time-independent expectation values for certain operators. 

These solutions may be readily generalised to conformally static spacetimes
where the conformal scale factor depends only on the time coordinate. 
This class includes the Robertson-Walker spacetimes with curved 
spatial sections.

Expectation values of operator products are given in the {\schr} representation
by
\be
  \langle 0 | \, O(\varphi, \pi) \, | 0 \rangle \:=\: \int {\cal
    D}\phi \: \Psi^\ast_0 \: O(\phi, -i \frac{\delta}{\delta \phi}) \:
  \Psi_0
\ee
where $\pi$ is the momentum conjugate to $\varphi$ in the canonical
formalism. Simple examples include
\begin{eqnarray}
  \langle 0 |\varphi(\underline{x}) \varphi(\underline{y})| 0 \rangle &=& 
  \half
  \Delta_{\cal R}(\underline{x},\underline{y};t)\\
  \langle 0 | \, \pi(\underline{x}) \, \pi(\underline{y}) \, | 0
  \rangle &=&
  \half \sqrt{h_x h_y} \Bigl\{
  \:G_{\cal R}(\underline{x},\underline{y};t) +
   \nonumber \\ &&
   \hspace*{-1.5cm}\int\! d^d\underline{u}
  \int\! d^d\underline{v}\, \sqrt{h_u h_v} \,
  G_{\cal I}(\underline{x},\underline{u};t)
  \Delta_{\cal R}(\underline{u},\underline{v};t)
  G_{\cal I}(\underline{v},\underline{y};t)
  \Bigr\}
\end{eqnarray}
and
\begin{eqnarray}
  \langle 0 | \, [\varphi(\underline{x}) , \pi(\underline{y})] \, | 0
  \rangle &=& i \delta^d(\underline{x}-\underline{y})\\
  \langle 0 | \, \{\varphi(\underline{x}) , \pi(\underline{y})\} \, | 0
  \rangle &=&- \int d^d\underline{u} \, \sqrt{h_x h_u} \,
  G_{\cal I}(\underline{x},\underline{u};t)
  \Delta_{\cal R}(\underline{u},\underline{y};t)
\end{eqnarray}
where $\Delta_{\cal R}$ is the inverse of the real part of the 
kernel $G_{\cal R}$. 

The expectation value of the canonical energy-momentum tensor
\begin{equation}
  T_{\mu \nu}(x) = (\partial_\mu \phi)(\partial_\nu \phi)
  -\frac{1}{2} g_{\mu \nu}
  [g^{\rho \sigma}(\partial_\rho \phi)(\partial_\sigma \phi) -
  (m^2 + \xi R)\phi^2]
\label{eq:emtensor}
\end{equation}
can be written in terms of the kernel and its inverse if we point-split
before calculating the expectation value, the coincidence limit being taken 
at the end of the calculation. 
In particular the expectation value of the `energy' component is
\begin{equation}
\langle0| T_{00}(x)|0\rangle = 
\lim_{\ux \rightarrow \uy}\: \langle0| T_{00}(\ux,\uy;t)|0\rangle
\label{eq:emtensorsplit}
\end{equation}
where 
\begin{eqnarray}
  \langle0| T_{00}(\ux,\uy;t) |0\rangle&=&
  -\frac{g_{00}}{2} \biggl\{ \Bigl(\frac{g_{00}}{g}\Bigr)
  \langle0| \pi(\ux) \pi(\uy)|0\rangle +\Bigl[
    g^{ij} \frac{\partial^2}{\partial x^i \partial y^j} -m^2\Bigr]
    \langle0|\phi(\ux) \phi(\uy) |0\rangle\biggr\} \nonumber \\
&=&
{g_{00}\over4} \biggl\{G_{\cal R}(\ux,\uy;t) - \Bigl[
 g^{ij} \frac{\partial^2}{\partial x^i \partial y^j} -m^2\Bigr]
\D_{\cal R}(\ux,\uy;t)  \nonumber\\
\label{eq:energytensor}
&&\hspace*{1cm}+\int d^d\uu \int d^d\uv \sqrt{h_x h_u}
G_{\cal I}(\ux,\uu;t)\D_{\cal R}(\uu,\uv;t)G_{\cal I}(\uv,\uy;t) \biggr\}
\end{eqnarray}

%%%%%%%%%%%%%%%%%%%%%%%%%%%%
\section{The Milne Universe}
%%%%%%%%%%%%%%%%%%%%%%%%%%%%

Our first example is a dynamic spacetime of Robertson-Walker type.
The Milne universe is a two-dimensional spacetime which begins at
an initial point and expands indefinitely. Quantum field theory in this
spacetime has been previously studied in 
\cite{disessa74, sommerfield74,gromes74}. 

The metric is 
\be 
ds^2 = dz^2 - a^2(z) d\t^2
\ee
where $z$ is the time coordinate ($z>0$) and $\t$ is the space coordinate
($-\infty<\t<\infty$). The scale factor for the Milne universe 
is $a(z) = z$.

With a rescaling of the time coordinate, it can be rewritten in manifestly
conformally flat form:
\be
ds^2 = C(\eta) (d\eta^2 - d\t^2)
\ee
where $\eta = \ln z$ and $C(\eta) = e^{2\eta}$.

\begin{figure}[htb]
\centerline{\epsfxsize=3.5in \epsfbox{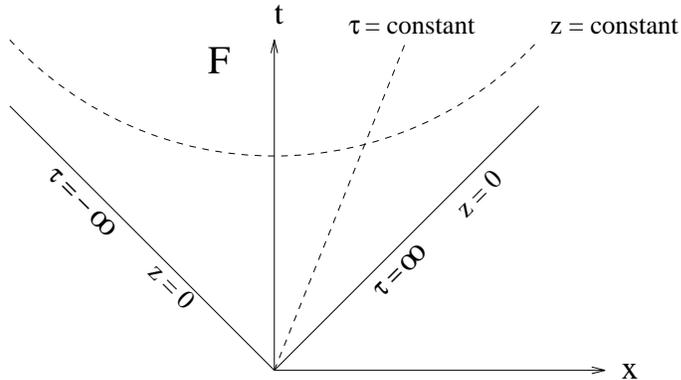}}
\caption{Milne patch of Minkowski spacetime.}
\label{fig:future}
\end{figure}

A further coordinate transformation, with $t = z\cosh \t$ and $x = z\sinh \t$,
brings the metric to the form
\be
ds^2 = dt^2 - dx^2
\ee
where the coordinates are restricted to the range $0<t<\infty$
and $-\infty<x<\infty$. In this form, it is clear that the Milne universe is
simply the patch of Minkowski spacetime lying in the future light cone
of the origin (see Fig~(\ref{fig:future})). This will be exploited in 
section 5.

The Milne universe is geodesically complete in the sense that it admits
a foliation where each spacelike hypersurface is intersected exactly
once by a semi-infinite timelike geodesic which does not
intersect the boundary\footnote{This is not true for null geodesics. 
In consequence, the conclusions of this section may not necessarily 
all be true for zero mass fields.} except at the special point at the origin.
A suitable foliation in which to set up the {\schr} formalism is 
shown in Fig~(\ref{fig:future})
where we choose the Cauchy hypersurfaces $\S$ to be the lines of constant $z$,
and consider evolution in the time coordinate $z$. 
The {\schr} equation for a minimally coupled massive scalar field is
\be
i {\pl\Psi\over\pl z} = {1\over2}\int_{-\infty}^\infty d\t~{1\over z}  \biggl[
-{\d^2\over\d\phi(\t)^2} +  \bigl[\pl_{\t} \phi(\t)\bigr]^2 
+ z^2 m^2 \phi(\t)^2 \biggr] \Psi
\label{eq:milne2d}
\ee 
where $\Psi[\phi,z]$ is a functional of the field eigenvalues $\phi(\t)$
on the equal-$z$ hypersurfaces. It may be solved as usual, giving
\be
\Psi[\phi,z] = N_0(z) \exp\left\{ -{1\over2} \int_{-\infty}^{\infty} d\t
\int_{-\infty}^{\infty} d\t' ~z^2 
\phi(\t) G(\t,\t'; z) \phi(\t')\right\}
\ee
where
\be
{d\over dz}\ln N_0(z) = -{i\over2} \int_{-\infty}^{\infty}d\t\:z\: G(\t,\t;z)
\ee
The kernel is
\begin{eqnarray}
G(\t,\t';z) &=& \int_{-\infty}^{\infty} {dk\over 2\p}
e^{ik(\t-\t')} \widetilde{G}(k;z) \\
& & \nonumber \\
\label{eq:milnekernel}
\widetilde{G}(k;z)&=& -{i\over z}{\pl\over\pl z} \ln \tpsi^\ast(z,k)
\end{eqnarray}
where $\tpsi(z,k)$ is a solution of the Fourier transformed wave equation
\be
\biggl({1\over z} \pl_z(z\pl_z) + m^2 +{k^2\over z^2}\biggr) \tpsi(z,k) = 0
\ee
The general solution is a linear combination of Hankel functions 
of imaginary order (see \cite{magnus66, grad80} for the required
properties of Hankel and Bessel functions), i.e. 
\be
\tpsi(z,k) = a(k)\,e^{-\frac{\pi k}{2}}\,H_{ik}^1(mz) + 
b(k)\,e^{\frac{\pi k}{2}}\,H_{ik}^2(mz)
\label{eq:milnehankels}
\ee
Since the kernel depends only on the logarithm of $\tpsi(z,k)$, only
the ratio of the coefficient functions $a(k)$ and $b(k)$ survives as a 
one-parameter ambiguity in the vacuum wave functional. 
To fix this, we need to choose a suitable boundary condition.

In the cosmological models considered in \cite{long96}, the spacetime had 
asymptotically Minkowski  regions and the boundary condition was specified
by choosing a vacuum wave functional that reproduced the standard
Minkowski vacuum in the asymptotic limit. This is achieved by
picking solutions of the wave equation which are positive frequency 
with respect to the usual Minkowski time coordinate. 
In the Milne universe, we have no analogous asymptotic region.
However, we can still require that $\tpsi(z,k)$ is a positive frequency
solution (more precisely, a sum of positive frequency solutions) with
respect to the proper time $z$ of comoving observers in the expanding
universe. Using a well-known integral representation of the Hankel functions
we may rewrite eq.(\ref{eq:milnehankels}) as 
\be
\tpsi(z,k) = 
-\frac{ia(k)}{\pi}\,\int_{-\infty}^\infty dt e^{iz \cosh t -ikt}  
+ \frac{ib(k)}{\pi}\,\int_{-\infty}^\infty dt e^{-iz \cosh t -ikt}   
\ee
So, remembering that for a comoving observer ($\t = {\rm const}$),
$z$ is simply proportional to $t$, we restrict $\tpsi(z,k)$ to be positive
frequency in the above sense by choosing $a=0$.
The vacuum wave functional is therefore specified by the kernel 
(\ref{eq:milnekernel}) with $\tpsi(z,k) = H_{ik}^2(mz)$.

To investigate the properties of this vacuum state, we evaluate first
the two-point Wightman Green function then the vacuum expectation
value of the energy-momentum tensor.
The two-point function evaluated at equal $z$-time is simply
\be
\langle 0|\varphi(\t) ~\varphi(\t')|0\rangle
~=~ {1\over2} \D_{\cal R}(\t,\t';z)
\ee
where $\D_{\cal R}(\t,\t';z)$ is the inverse of the real part of the kernel, 
$G_{\cal R}(\t,\t';z)$. Defining the Fourier transform by
\be
\D_{\cal R}(\t,\t';z) = \int_{-\infty}^{\infty} {dk\over2\p} 
\widetilde\D_{\cal R}(k;z) e^{ik(\t-\t')}
\ee
we find 
\be
\widetilde \D_{\cal R}(k;z) = - {2i\over z} |\tpsi(z,k)|^2 W^{-1}[\tpsi^*(z,k),
\tpsi(z,k)]
\ee
For the general solution (\ref{eq:milnehankels}), we have
\begin{eqnarray}
|\tpsi(z,k)|^2 &=& (|a|^2 + |b|^2) 
H_{ik}^1(mz) \,H_{ik}^2(mz) \nonumber \\
&+& a\,b^*\, e^{-\p k}\, H_{ik}^1(mz)\, H_{ik}^1(mz)
+ a^*\,b\,e^{\p k}\, H_{ik}^2(mz)\, H_{ik}^2(mz)
\end{eqnarray}
while the Wronskian is 
\begin{eqnarray}
W[\tpsi^*(z,k), \tpsi(z,k)] &=&(|a|^2  - |b|^2)
W[H_{ik}^2(mz), H_{ik}^1(mz)]  \nonumber \\
&=& {4 i\over \p z} (|a|^2 - |b|^2)
\end{eqnarray}
So, for the chosen comoving vacuum ($a = 0$), we find
\be
\langle 0|\varphi(\t)~\varphi(\t')|0\rangle_{\rm COM}
= {\p\over4} \int_{-\infty}^{\infty} {dk\over 2\p} e^{ik(\t-\t')}
H_{ik}^1(mz) H_{ik}^2(mz)
= {1\over2\p} K_0(m\s)
\ee
where $\s$ is the geodesic interval along the equal-$z$ hypersurface of the
foliation, viz.
\be
\s = 2z \sinh\Bigl({\t-\t'\over2}\Bigr)
\label{eq:sigmaf}
\ee
Expressed in Minkowski coordinates $(t,x)$, 
\be
\s^2 = (x-x')^2 - (t-t')^2
\label{eq:sigmamink}
\ee
The details of the calculation are given in section 5.

We see, therefore, that the two-point function in the comoving vacuum
in the Milne universe is identical to the corresponding Green function
in the complete Minkowski spacetime. This is not too surprising since
we have used the same boundary condition in choosing the vacuum state,
although it is less obvious that the Green function should be insensitive
to the boundary, recalling that the Milne universe is simply the
patch of Minkowski spacetime in the future light cone of the origin.
This is assured by the property that the Milne patch admits
a foliation for which the spacelike hypersurfaces are complete Cauchy
surfaces for the full Minkowski manifold. This property is not shared by the 
other related example considered in this paper, the Rindler wedge (section 4).

As a second probe of the vacuum state, we may evaluate the expectation
value of the energy-momentum tensor, eq.(\ref{eq:emtensor}). The
`energy' component is expressed in terms of the expectation value 
with point-split argument, eq.(\ref{eq:energytensor})
\begin{eqnarray}
 \langle0| T_{zz}(\t,\t';z)|0\rangle&=&
{1\over4} \biggl\{G_{\cal R}(\t,\t';z) +\left[
  \frac{1}{z^2}
  \frac{\partial^2}{\partial \t \partial \t'} 
 +m^2\right]\:\D_{\cal R}(\t,\t';z)  +\nonumber\\
&&\hspace*{-0.8cm}z^2\int_{-\infty}^{\infty}\!\!\!d\t'' \!\! 
\int_{-\infty}^{\infty}\!\!\!d\t'''
G_{\cal I}(\t,\t'';z)\D_{\cal R}(\t'',\t''';z)G_{\cal I}(\t''',\t';z) \biggr\}
\end{eqnarray}
After some calculation (see appendix A for details), we find for the 
comoving vacuum, 
\be
\langle 0| T_{zz}(\t,\t';z)|0\rangle_{\rm COM} =
-\left(\frac{m}{2\pi \sigma}\right)\:K_1(m\sigma)
\ee
where $\sigma$ is the geodesic interval along the equal-z hypersurfaces
of the foliation. Again, this agrees with the point-split energy-momentum 
tensor VEV for Minkowski  spacetime, allowing for the coordinate 
transformation to the $(z,\t)$ coordinates.

These two results confirm that physical quantities calculated in the 
Milne universe with the particular choice of state we have called
the comoving vacuum are identical to those in Minkowski spacetime.
In particular, they show no dependence on the boundary.
However, other equally valid choices of vacuum state are possible
corresponding to different choices of the arbitrary ratio $a/b$ in 
eq.(\ref{eq:milnehankels}).
We now consider one of these, the so-called `conformal' vacuum.

The conformal vacuum\footnote{Notice that, 
as in \cite{long96}, we could equally 
well have formulated the {\schr} equation for evolution in the conformal 
time, i.e.~along the conformal Killing vector ${\pl\over\pl\eta}$. However, 
since $\eta$ is a function of $z$ only (recall $\eta= \ln z$), the 
foliations into $z={\rm const}$ and $\eta = {\rm const}$ surfaces are 
identical, so the {\schr} equations are related by a trivial change of 
variable. In contrast, the choice of vacuum 
state is made at the level of imposing a boundary condition on the kernel
equation. The comoving and `conformal' vacua are distinguished by the choice
of $\tpsi$ to be positive frequency with respect to the proper time $z$
of a comoving observer or (for massless fields) the conformal time $\eta$
respectively. This is a physical distinction unrelated to the foliation 
choice.}
is selected by requiring that in the massless limit,
where we are considering a conformal field theory on a conformally
flat spacetime, the wave equation solutions determining the kernel
should be positive frequency with respect to the conformal 
time $\eta$.

Yet another rewriting of the general solution (\ref{eq:milnehankels}) to the wave equation
gives
\be
\tpsi(z,k) = c(k) J_{i|k|}(mz) + d(k) J_{-i|k|}(mz)
\label{eq:milnebessels}
\ee
where the $J_{\pm i|k|}(mz)$ are Bessel functions of imaginary order.
In terms of the $a$ and $b$ coefficients of eq.(\ref{eq:milnehankels}), 
we have for $k>0$
\begin{eqnarray}
c(k) &=& \a(k)\,a(k) + \b(k)\,b(k) \\
d(k) &=& \b^*(k)\,a(k) +  \a^*(k)\,b(k)
\end{eqnarray}
where $\a(k)$ and $\b(k)$ are Bogoliubov coefficients:
\be
\a(k) = \frac{e^{\frac{\pi k}{2}}}{\sinh (\pi k)} \hspace{1.5cm} \b(k)
 = - \,\frac{e^{-\frac{\pi k}{2}}}{\sinh (\pi k)}
\ee
In the massless limit, $J_{-i|k|}(mz) \sim z^{-i|k|} = e^{-i|k|\eta}$,
so is positive frequency with respect to the conformal time.
The conformal vacuum is therefore specified by choosing
$c=0$, $d=1$ in eq.(\ref{eq:milnebessels}). In terms of the 
original coefficients, it is specified by choosing the ratio 
$a/b = e^{-\pi k}$. Clearly, it is simply another of the one-parameter 
family of possible vacua characteristic of dynamic spacetimes. 
Of course, the conformal vacuum also corresponds to
positive frequency behaviour with respect to the conformal time
in the limit of early times (small $z$), as well as vanishing mass.

To show that this is indeed physically distinct from the comoving vacuum,
we compare the expectation values of the energy-momentum tensor in the 
two states. Details of the calculations are given in appendix A.
We find that the difference between the expectation values in the comoving 
and conformal vacua \cite{bunch78} is 
\begin{eqnarray}
 &&\hspace*{-1cm}\langle 0| T_{zz}(z)|0\rangle_{\rm COM} - 
  \langle 0| T_{zz}(z)|0\rangle_{\rm CONF} = \frac{1}{\pi z^2}
  \int^\infty_0 \frac{dk\:k}{(e^{2 \pi k}-1)}  \nonumber \\
 &&+  \frac{m^2}{8} \int^\infty_0 
  \frac{dk}{\sinh^2(\pi k)} \biggl\{
  2 e^{-\pi k}\Bigl[J_{ik}(mz)J_{-ik}(mz)+ J_{ik+1}(mz)J_{-ik+1}(mz)\Bigr]
    \nonumber \\
  &&\hspace*{3.5cm} -J_{ik}(mz)J_{ik}(mz)-    J_{-ik}(mz)J_{-ik}(mz) + \nonumber \\
  && \hspace*{3.5cm} J_{ik-1}(mz)J_{ik+1}(mz) +J_{-ik-1}(mz)J_{-ik+1}(mz)
\biggr\} 
\end{eqnarray}
The first term dominates in the early time (small $z$) or small mass limits,
since all the other terms are of $O(z^0)$.  This term represents the
energy density of radiation at a temperature $(2 \pi z)^{-1}$, 
and shows that, in this limit, the comoving vacuum is an excited, 
thermal state with respect to the conformal vacuum. 

%%%%%%%%%%%%%%%%%%%%%%%%%%%
\section{Rindler Spacetime}
%%%%%%%%%%%%%%%%%%%%%%%%%%%

\begin{figure}[htb]
\centerline{\epsfxsize=2.5in \epsfbox{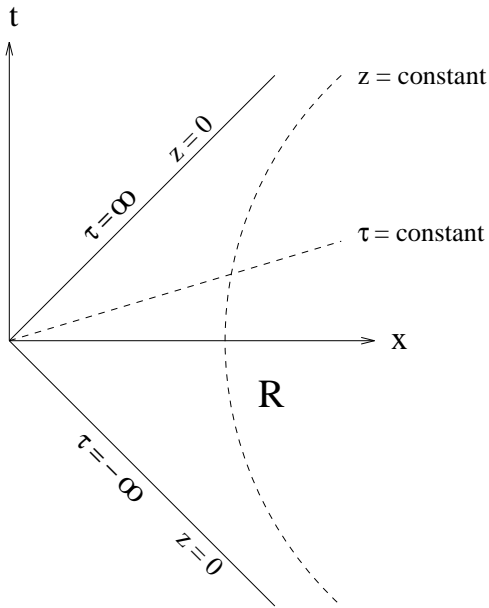}}
\caption{Rindler wedge of Minkowski spacetime.}
\label{fig:right}
\end{figure}

Rindler spacetime \cite{rindler66} is the static spacetime described 
by the two-dimensional metric
\be
ds^2 = z^2 d\t^2 - dz^2
\ee
with $-\infty<\t<\infty$ and $0<z<\infty$.
In coordinates which make the conformal flatness manifest,
\be
ds^2 = C(\eta) \bigl(d\t^2 - d\eta^2\bigr)
\ee
where $\eta = \ln z$ and $C(\eta) = e^{2\eta}$.
Like the Milne universe, Rindler spacetime is simply a patch of Minkowski
spacetime. To see this, make the coordinate transformation
$t = z\sinh\t$, ~$x = z\cosh\t$. In these coordinates, the metric is simply
\be
ds^2 = dt^2 - dx^2
\ee
where the range is restricted to $x>0$,~$|t|<x$. The spacetime is therefore
just the {\bf R} wedge in Fig~(\ref{fig:right}).

Quantum field theory in this spacetime has been widely studied using many
different formalisms (see for example \cite{fulling73, unruh76, sciama81}
in the canonical formalism and \cite{freese85, hill85, hill86} in the 
{\schr} formalism). We have little to add to this discussion so the presentation 
here is very brief. It is intended mainly to illustrate the importance of
boundary conditions in specifying the vacuum state and to contrast with
the results on foliation independence in the Rindler-Milne analysis
of Minkowski spacetime in section 5.

In order to apply the {\schr} formalism, we need to choose a foliation into a 
set of spacelike Cauchy hypersurfaces and consider evolution along
a timelike Killing vector field which is infinite in extent and in particular 
does not intersect the boundary. The Rindler wedge is globally hyperbolic
and thus geodesically complete and so admits such a foliation.

A suitable foliation is given by choosing the spacelike hypersurfaces
to be the lines $\t = {\rm const}$ and considering evolution along the 
Killing vectors $\pl/\pl\t$ as shown in 
Fig~(\ref{fig:right})\footnote{The evolution path
$z={\rm const}$ is the world line of a uniformly accelerating
particle with acceleration $1/z$ in Minkowski spacetime. This is 
the reason for the great interest in Rindler spacetime \cite{sciama81} 
in modelling the behaviour of accelerated systems or observers.}. 
The {\schr} equation is then
\begin{equation}
  i{\frac{\partial\Psi}{\partial \tau}} = 
  \half \int_0^\infty \! \! dz   \: z 
\Bigl\{ \!-\frac{\delta ^2}{\delta \phi(z) ^2}\,
 + [\partial_z \phi(z)]^2 \, + \, m^2\phi(z)^2\Bigr\} \Psi.
\end{equation}
To solve this, we must impose boundary conditions on the field $\phi(z)$.
A suitable choice is the Dirichlet condition $\phi = 0$ at $z=0$
(and, as usual, at spatial infinity, $z\rightarrow\infty$).
The vacuum wave functional is
\be
\Psi[\phi,\t] = N_0(\t) \exp\Bigl\{-\half \int_0^\infty dz \,
\int_0^\infty dz' \,  \phi(z)G(z,z')\phi(z')\Bigr\}
\ee
where $N_0(t) = \exp \{-\frac{i}{2} \int_0^\infty dz \,z G(z,z)\}$ 
and the kernel $G(z,z')$ is given by the general formula
for static spacetimes, in this case
\be
G(z,z') = \frac{1}{z z'} \int_0^\infty \frac{d\w}{2\pi}\, \w\:
\tpsi(\w,z)\:\tpsi^\ast(\w,z')
\ee
The functions $\tpsi(\w,z)$ are Fourier transforms with respect to $\t$
of solutions of the wave equation, viz.
\be
\Bigl(\frac{1}{z}\,\partial_z(z\,\partial_z) - \,m^2 + 
\frac{\w^2}{z^2} \Bigr)\,\tpsi(\w,z) = 0 
\ee
The boundary condition on $\phi(z)$ is respected automatically if
we choose $\tpsi(\w,z)$ such that $\tpsi = 0$ at $z=0$. A suitable set,
satisfying the orthonormality and completeness conditions
\begin{eqnarray}
  \int_0^\infty  \frac{dz}{2\pi} 
    \,\frac{1}{z}\,  \tpsistar(\w,z)\,\tpsi(\nu,z)\,
  &=&\delta(\w - \nu) \\
  \int_0^\infty  \frac{d\w}{2\pi} 
  \tpsistar(\w,z)\,\tpsi(\w,z')\,
  &=&z\: \delta(z-z')
\label{eq:comporthLR}
\end{eqnarray}
is\footnote{These results are immediate consequences of the 
Kontorovich-Lebedev transform
 \begin{eqnarray}
  g(y) &=& \int_0^\infty dx\: f(x) K_{ix}(y)\nonumber \\
  f(x) &=& 2\pi^{-2}\,x\,\sinh(\pi x) 
  \,\int_0^\infty dy\:y^{-1}\: g(y)\: K_{ix}(y) \nonumber
\end{eqnarray}
}
\be
\tpsi(\w,z) =  2\sqrt{\frac{\w\,\sinh(\pi \w)}{\pi}}
  \:K_{i\w}(mz).
  \label{eq:wavesolutionLR}
\ee

This specifies the vacuum state in Rindler spacetime subject to the
given boundary condition.  It is the ground state
with respect to the energy associated with the chosen time evolution.
It is unique in the same sense as is the usual vacuum in Minkowski spacetime. 
Of course, a different foliation satisfying the above criteria would yield 
a formally different expression for the vacuum wave functional, but
all physical quantities derived from it would be identical.
(The question of foliation independence is discussed in section 5.)

An alternative representation of the wave functional can be given
in terms of the transforms $\tphi(\w)$ of the field 
eigenvalues $\phi(z)$,
\begin{equation}
  \phi(z) =  \int_0^\infty \frac{d\w}{2\pi}\: 2\:
  \sqrt{\frac{\w\,\sinh(\pi \w)}{\pi}}  K_{i\w}(mz) \tphi(\w).
\end{equation}
As a functional of $\tphi$, the $\t$-independent part of the
vacuum wave functional is simply
\begin{equation}
  \label{eq:wavefnalmomentumrind}
  \Psi[\tphi(\w)] = 
  \exp\Bigl\{-\half  \int_{0}^{\infty} \frac{d\w}{(2\pi)} 
  \:\w\: \: |\tphi(\w)|^2\Bigr\}.
\end{equation}
Excited states can be built as described in \cite{long96} by 
acting successively on $\Psi[\tphi]$ with the creation operators
\be
a^\dagger(\w) = \int_0^\infty dz\,\tpsi(\w,z) \Bigl[
\frac{\w}{z} \phi(z) - \frac{\delta}{\delta \phi(z)}\Bigr]
\ee

To understand better the nature of this vacuum state, we again evaluate
the Wightman function and the energy-momentum tensor.
The Wightman function is simply the inverse kernel. Clearly, we have
\begin{eqnarray}
\Delta(z,z') &=& \int_0^\infty \frac{d\w}{2\pi} \:
\frac{1}{\w}\: \tpsi(\w,z)\: \tpsi^\ast(\w,z') \\
&=& \frac{4}{\pi}  \int_0^\infty \frac{d\w}{2\pi} \: 
\sinh(\pi \w)  K_{i\w}(mz)K_{i\w}(mz')
\end{eqnarray}
and evaluating the integral over $\w$ we find
\begin{equation}
  \label{eq:vevphiphiR}
  \langle 0 | \varphi(z) \varphi(z') | 0 \rangle_{\rm RIND}
  = \frac{1}{2\pi}    K_0(m|z-z'|) - 
  \frac{1}{\pi} \int_0^\infty \frac{dv}{\pi^2+v^2}\:
  K_0(m Z)
\end{equation}
where $Z^2 = z^2 +z'^2 +2zz'\cosh v$. The first term is simply the 
usual translation invariant Minkowski result. 
(Note that the geodesic interval $\s^2 = (t-t')^2 - (x-x')^2$
for points with equal $\t$ is simply $\s^2 = -(z-z')^2$).
The second term shows a dependence on the absolute
position and reflects the sensitivity to the boundary.
This should be contrasted with the corresponding result in the Milne universe.
The foliation hypersurfaces for the Rindler patch are not complete Cauchy
hypersurfaces for the full Minkowski spacetime, so there is no reason
to expect translation invariance in the Wightman function.

The energy-momentum tensor expectation values are computed as
usual from the kernel and its inverse. After some calculation 
(see appendix B) we find
\begin{eqnarray}
  &&\langle 0|T_{\t\t}(z,z')|0\rangle_{\rm RIND} 
  = -z^2\Bigl(\frac{m}{2\pi |z-z'|}\Bigr) K_1(m|z-z'|)
  +\int_0^\infty \frac{dv \:z^2}{2\pi (v^2+\pi^2)}\times \nonumber \\
  &&\hspace*{2cm}\mbox{$\biggl[\Bigl(\frac{P(v)}{z z'}-Q(v)m^2\Bigr)K_0(mZ) +
\frac{m}{Z}\Bigl(2 + \cosh v -2 Q(v)\Bigr)K_1(mZ)
\biggr]$}
\end{eqnarray}
The first term is exactly the same (up to a factor of $g_{\tau \tau}$)
as the usual Minkowski result and depends only on the geodesic 
interval between the points. The second term, however, is not 
translation invariant and shows an explicit position dependence.

The energy density appropriate to evolution along the Killing vectors
$\pl/\pl\t$ is therefore position dependent and sensitive to the boundary.
However, if instead we calculate the expectation value of  the corresponding
Hamiltonian, given by
\be
\langle 0| H_R|0\rangle_{\rm RIND} = 
\int_0^\infty dz \sqrt{-g} g^{00} \langle 0|T_{zz}(\t,z)|0\rangle_{\rm RIND}
\ee
we simply find the usual Minkowski-like sum of zero-point energies, viz.
\be
\langle 0|H_R|0\rangle_{\rm RIND} = {1\over2} \int_0^{\infty} d\w~ \w ~\d(0)
\ee

The Rindler vacuum therefore shares most of the properties of the  
familiar Minkowski vacuum. It is the ground state with respect to the
energy associated with the Hamiltonian generating the time evolution
along the vector field $\pl/\pl\t$. A simple spectrum of excited states
is generated by the creation operators $a^{\dagger}(\w)$.
However, the lack of translation invariance in Rindler spacetime does
affect the vacuum, showing up both in the Wightman function
and in the explicit position dependence, or boundary sensitivity, of the 
local energy density.
\vskip0.3cm

Finally, we should make some remarks about observer dependence
in the interpretation of this Rindler vacuum state.

In Minkowski spacetime, the Unruh effect implies that the vacuum state
appears simple only to the class of inertial observers, whereas uniformly 
accelerated observers will experience a universal temperature effect
\cite{unruh76,bell83}.

In Rindler spacetime, the r\^ole of preferred observers is taken by those 
following the timelike Killing vector fields $\pl/\pl\t$. These observers will be
the analogues of the inertial observers in Minkowski spacetime and will 
perceive the Rindler vacuum to be a simple vacuum state.
Other observers are accelerated relative to this class and will therefore 
experience an Unruh effect, perceiving the Rindler vacuum to be an
excited state. For example, we expect observers following the 
Minkowski time evolution vectors $\pl/\pl t$ to experience a universal,
position-dependent temperature effect ($T = 1/2\pi z$), with the
temperature increasing as the boundary is approached.
This behaviour is in complete contrast to that of observers following
the geodesically complete vector fields $\pl/\pl\t$, which are infinite in
extent and never intersect the boundary.

%%%%%%%%%%%%%%%%%%%%%%%%%%%%%%%
\section{Rindler-Milne Foliation of Minkowski Spacetime}
%%%%%%%%%%%%%%%%%%%%%%%%%%%%%%%

This final example is designed to illustrate the foliation independence
of physical quantities for quantum field theories in the same spacetime
with the same boundary conditions. 

In general, the foliation is
specified by the deformation vector $N^\m(x)$ (which incorporates the 
lapse and shift functions $N$ and $N^i$).
The foliation determines the representation of operators in terms of the fields
$\varphi$ and conjugate momenta $\pi$, so that both the operators and the
wave functionals depend on $N^\m$. 
Foliation independence of physical quantities would then be expressed as a 
functional Ward identity with respect to $N^\m$.
For example, for the physical VEV of a renormalisation group invariant
operator $O(\pi,\varphi;N^\m)$, we would have a Ward identity of the form
\be
{\d\over\d N^\m} \int {\cal D}\phi~
\Psi^*[\phi,s;N^\m] ~O\biggl(-i{\d\over\d\phi},\phi;N^\m\biggr)~
\Psi[\phi,s;N^\m]
~=~ 0
\ee
This encodes the invariance of the VEV under infinitesimal changes
of the foliation hypersurfaces, although the wave functional itself
is of course foliation dependent. 

\begin{figure}[htb]
\centerline{\epsfxsize=2.5in \epsfbox{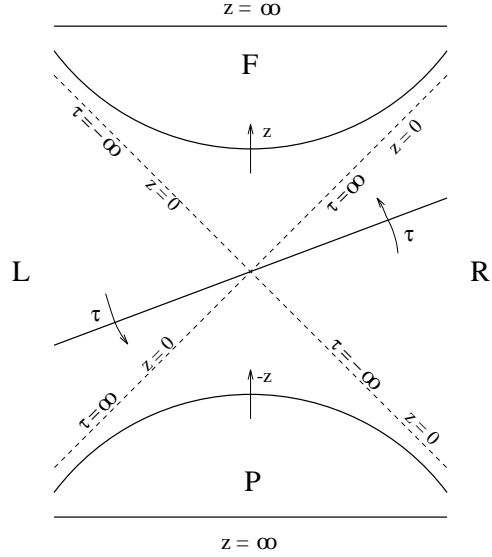}}
\caption{Rindler--Milne evolution surfaces in Minkowski spacetime.}
\label{fig:evln}
\end{figure}

In this section, however, we consider `large' changes of foliation.
The example we choose is ordinary, $(d+1)$ dimensional Minkowski 
spacetime and we consider two foliations, first the standard one with
hypersurfaces $t = {\rm const}$ and second a `Rindler-Milne'
foliation where the spacetime is split into sections {\bf P, L+R, F}
and the spacelike hypersurfaces are as shown in Fig~(\ref{fig:evln}).

%%%%%%%%%%%%%%%%%%%%%%%%%%%%%%%
\subsection{Minkowski foliation}

The results of the standard Minkowski foliation \cite{long96} are well 
known and we 
simply quote them. The vacuum wave functional, which satisfies the 
{\schr} equation
\begin{equation}
i{\frac {\pl\Psi[\phi,t]}{\pl t}}
 = \frac{1}{2} \int{d^d\ux}~
\Bigl\{ -\frac{\delta^2}{\delta \phi^2} ~-~
\eta^{ij}(\pl_i\phi) (\pl_j\phi)
~+~ m^2\phi^2\Bigr\} \Psi[\phi,t] \label{eq:swfemink}
\end{equation}
is
\be
\Psi_0[\phi,t] = N_0(t) \
\exp\left\{-\frac{1}{2}\int{d^d\ux}\int{d^d\uy} \
\phi(\ux)\,G(\ux,\uy)\,\phi(\uy)\right\}
\label{eq:psi0mink}
\ee
where the kernel is
\be
 G(\ux,\uy)=\int 
\frac{d^d\uk}{(2\pi)^d}\:\sqrt{\uk^2 + m^2}
 \: {e^{i\uk.(\ux-\uy)}} 
\ee

The inverse kernel gives the Wightman function on a $t={\rm const}$ hypersurface,
viz.
\be
\langle 0|\varphi(\ux)~\varphi(\uy)|0\rangle_{\rm MINK} =
{1\over2\pi}\left({\frac{m}{2\pi|\ux-\uy|}}\right)^{{d-1\over2}}
  K_{{d-1\over2}}(m|\ux-\uy|)
\label{eq:invkernel}
\ee
Notice that due to the manifest translation invariance, the Green function
depends only on the distance $|\ux -\uy|$ separating the points. 

The (unrenormalised) VEV of the energy--momentum tensor is just the 
usual sum of zero-point energies,
\be
\langle 0|T_{\m\n}(x)|0\rangle_{\rm MINK} = g_{\m\n} \frac{1}{2}
\int \frac{d^d\uk}{(2\pi)^d}\:\w(\uk)
\ee
where $\w^2(\uk) = \uk^2+m^2$. For later comparison the VEV of the 
energy component of the point--split energy--momentum tensor is
\begin{equation}
  \label{eq:toomink}
  \langle 0|T_{00}(\ux,\uy)|0\rangle_{\rm MINK} = -\left(
    \frac{m}{2 \pi |\ux -\uy|}\right)^{\frac{d+1}{2}}\:
  K_{\frac{d+1}{2}}(m |\ux -\uy|)
\end{equation}

\subsection{Rindler-Milne foliation}

We now compare these results with those for the Rindler-Milne foliation.
To set this up, we split Minkowski spacetime into the four wedges
shown in Fig~(\ref{fig:evln}) and introduce coordinates $(\t,z,x^a)$ in each wedge as
follows:
\begin{equation}
\begin{array}{llcr}
x^1=z\cosh \tau \;\;\;\;\;\;\; &t=z\sinh \tau&\hspace{1cm} &x^1,t\in R\\
x^1=-z\cosh \tau \;\;\;\;\;\; &t=-z\sinh \tau& &x^1,t\in L\\
x^1=z\sinh \tau \;\;\;\;\;\;\; &t=z\cosh \tau& &x^1,t\in F\\
x^1=-z\sinh \tau \;\;\;\;\;\;\; &t=-z\cosh \tau& &x^1,t\in P.
\label{eq:MERcoordinatetransformations}
\end{array}
\end{equation}
$x^a$ ($a=2,\dots,d$) are retained as Minkowski coordinates.

\vskip0.5cm
\noindent{\bf F and P patches}
\vskip0.2cm
In the {\bf F} and {\bf P} patches, the metric is $ds^2 = dz^2 - z^2 d\t^2 - (dx^a)^2$
so appears dynamic in these coordinates. The spacelike hypersurfaces are 
chosen to be $z = {\rm const}$ and we consider evolution along 
$\pl/\pl\t$. These hypersurfaces are complete Cauchy surfaces for the whole 
of the Minkowski spacetime.

The analysis is precisely as in section 3, except that here we are working in
$(d+1)$ dimensions. The {\schr} equation is just the generalisation of 
eq.(\ref{eq:milne2d}) and the vacuum wave functional is
\begin{equation}
  \Psi[\phi(\tau, x^a),z] = N_0(z) 
  \exp \Bigl\{-\half
  \int d^d\underline{x} \: \int d^d\underline{y}
  \, z^2 \phi(\underline{x})G(\underline{x}, \underline{y};z)
  \phi(\underline{y})\Bigr\}
\end{equation}
with kernel
\begin{eqnarray}
  G(\underline{x},\underline{y};z)&=&
  \int_{-\infty}^{\infty} \frac{d\w}{(2\pi)} \: 
  \int \frac{d^{d-1}k_a}{(2\pi)^{d-1}}\:
  e^{i\w(\tau-\tau')}\,e^{ik_a(x^a-y^a)}\widetilde{G}(\uk;z) \\
  \label{eq:milnekernel2}
  \widetilde{G}(\uk;z) &=& -  \frac{i}{z} \frac{\partial}{\partial z}\,
  \ln\,\tpsi^\ast(z,\w, k_a)
\end{eqnarray}
Choosing boundary conditions on the wave equation solution $\tpsi(z,\w,k_a)$
as in section 3 so that it is a superposition of eigenfunctions which are 
positive frequency with respect to Minkowski time $t$, we have
\begin{equation}
  \tpsi(\w, k_a,z) = H^2_{i\w}(qz)
\end{equation}
where $q^2 = k_a^2 + m^2$. This resolves the one-parameter
ambiguity of vacuum states in this foliation.

The two point function evaluated at equal $z$ times in this vacuum
state is given in term of the inverse kernel, which can be shown 
(as in section 3) to be
\begin{equation}
  \langle 0 | \, \varphi(z,\t,x^a) \, \varphi(z,\t',x^a) 
  \, | 0 \rangle = \frac{\pi}{4}\:\int_{-\infty}^\infty \frac{d\w}{(2\pi)} \: 
  \int \frac{d^{d-1}k_a}{(2\pi)^{d-1}}\:
  e^{i\w(\tau-\tau')}\:H^1_{i\w}(qz)\:H^2_{i\w}(qz).
\end{equation}
where we have only considered points separated  in the $x^1$ direction. 
Rewriting the Hankel functions in
terms of modified Bessel functions and performing the $\w$ integral 
gives
\[
  \langle 0 | \, \varphi(z,\t,x^a) \, \varphi(z,\t',x^a) 
  \, | 0 \rangle =
  \frac{2}{2^d \pi^{\frac{d+1}{2}} \Gamma(\frac{d-1}{2})}
  \int_0^\infty dk \:k^{d-2} K_0(m \sigma)
\]
with $\sigma = 2 z \sinh(\frac{\tau -\tau'}{2})$. The remaining integral can
be performed by a Mellin transform to give the final expression for the 
two point function in this vacuum,
\begin{equation}
  \label{eq:phiphiF}
  \langle 0 | \, \varphi(z,\t,x^a) \, \phi(z,\t',x^a) 
  \, | 0 \rangle =     \frac{1}{2\pi}
    \:\left(\frac{m}{2\pi\sigma}
  \right)^{\frac{d-1}{2}}
  \:K_{\frac{d-1}{2}}(m\sigma)
\end{equation}

A similar calculation (see appendix A for details) gives 
the VEV of $T_{zz}$ with point-split arguments,
\begin{equation}
  \langle 0| T_{zz}(\t,\t';z)|0\rangle =
-\left(\frac{m}{2\pi \sigma}\right)^{\frac{d+1}{2}}\:
K_{\frac{d+1}{2}}(m\sigma)
\end{equation}
Comparing with the equivalent Minkowski spacetime
results, eqs.(\ref{eq:invkernel}) and (\ref{eq:toomink}),
and noticing that $\sigma$ is simply the geodesic interval between the 
points $(\t,x^a,z)$ and $(\t',x^a,z)$ as explained in section 3,
we see that as expected they are identical.

\vskip0.5cm
\noindent{\bf L-R patch}
\vskip0.2cm
As already observed in section 4 where we studied the single Rindler
wedge {\bf R}, the hypersurfaces $\tau={\rm const}$ in this patch alone
are not Cauchy complete in the extended Minkowski spacetime.
To find such surfaces, which are necessary to have a correct foliation 
of the spacetime (i.e.~respecting the global hyperbolicity and
geodesic completeness), we have to treat the {\bf L} and {\bf R} wedges
together. The metric for both patches is $ds^2 = z^2 d\tau^2 - dz^2 -(dx^a)^2$.
The Cauchy hypersurfaces are then the surfaces $\tau={\rm const}$
across both patches taken together, as shown in Fig(\ref{fig:evln}), 
and we consider evolution in $\pl/\pl\t$ as shown.

The {\schr} equation is then
\begin{equation}
i{\frac {\partial\Psi}{\partial \tau}} = 
\half \int_{\Sigma} d^d\underline{x} \: z 
\Bigl\{ \!-\frac{\delta ^2}{\delta \phi(\underline{x}) ^2}\,
 + [\partial_z \phi(\underline{x})]^2\,+\,  [\partial_a \phi(\underline{x})]^2\, 
+ \, m^2\phi(\underline{x})^2\Bigr\} \Psi 
\end{equation}
where the integral over the hypersurface is given by
\begin{equation}
  \label{eq:xintLR}
  \int_{\Sigma} d^d\underline{x} =  \int d^{d-1}x^a\: \left[
    \int^0_\infty dz \:\Theta_x(L) + \int_0^\infty dz \:\Theta_x(R)\right]
\end{equation}
$\Theta_x(L)$ is a theta function which is $0$ when x is in the 
{\bf R} region and is  1 when x is in the {\bf L} region. The vacuum 
wave functional solution is 
\begin{equation}
  \Psi[\phi(z,x^a),\tau] = N_0(\tau) 
  \exp \Bigl\{-\frac{1}{2} \int_\Sigma d^d\underline{x} \: 
  \int_\Sigma d^d\underline{y}\:
  \phi(\underline{x})G(\underline{x}, \underline{y})\phi(\underline{y})\Bigr\}
\end{equation}
with kernel\footnote{We use the notation $\ux = (z,x^a)$ and 
$\uy = (z',y^a)$ for space coordinates and $\uk = (\omega,k_a)$ and
$\up = (\nu,p_a)$ for momenta.}
\begin{equation}
  \label{eq:kernelLandR}
  G(\underline{x}, \underline{y}) = \frac{1}{z\,z'}
  \:\int_{-\infty}^\infty \frac{d\omega}{(2\pi)} \: 
  \int \frac{d^{d-1}k_a}{(2\pi)^{d-1}}\:
  \omega\,\tpsistar_{\uk}(\omega,\underline{x})\,\tpsi_{\uk}(\omega,
  \underline{y})
\end{equation}
To construct $G(\ux,\uy)$ we need a complete orthonormal set of 
eigenfunctions of the wave equation
\be
\Bigl(\frac{1}{z}\,\partial_z(z\,\partial_z) + \partial_{x^a}^2
 - \,m^2 + \frac{\omega^2}{z^2} \Bigr)\,\tpsi(\omega,z,x^a) = 0 
\ee
The unique set\footnote{We have already used the $\t$ independence
of the metric to show that the kernel is a function of $\ux, \uy$ only and 
Fourier transformed with respect to $\t$ to find solutions $\tpsi(\w,z,x^a)$
of the wave equation. As in Minkowski spacetime, this carries
an implicit definition of the positive frequency convention.} consistent 
with the boundary condition that the field eigenvalues $\phi(z,x^a)$ in the 
wave functional tend to zero at spatial infinity is found to be 
\begin{equation}
  \tpsi_{\uk}(\omega,x,x^a) = 
  \sqrt{\frac{2\w}{\pi}}
  \,e^{i k_a x^a}\:K_{i\w}(qz)\Bigl[e^{\frac{\pi \w}{2}} \Theta_x(R) + 
  e^{-{\frac{\pi \w}{2}}} \Theta_x(L)\Bigr]
  \label{eq:waveLandR}
\end{equation}
with $q^2 = (k_a^2 + m^2)$. These functions satisfy
\begin{eqnarray}
  \int_{\Sigma} \frac{d^d\underline{x}}{(2\pi)^d} \,\frac{1}{z}
  \tpsistar_{\uk}(\omega,\underline{x})\,\tpsi_{\up}(\omega,\underline{x})\,
  &=&\delta(\omega - \nu) \delta^{d-1}(k_a - p_a) \\
\label{eq:comporthLandR}
  \:\int_{-\infty}^\infty \frac{d\omega}{(2\pi)} \: 
  \int \frac{d^{d-1}k_a}{(2\pi)^{d-1}}\:
  \tpsistar_{\uk}(\omega, \ux)\,\tpsi_{\uk}(\omega,\uy)\,
  &=&z  \delta^{d-1} (x^a-y^a) \delta(z-z')\nonumber \\
  &&\:\times\: \bigl[ \Theta_x(R)\Theta_y(R) - \Theta_x(L)\Theta_y(L)\Bigr]\nonumber
\end{eqnarray}
The two point function evaluated at equal $\tau$ times is given 
in terms of the inverse kernel
\begin{equation}
  \Delta(\underline{x},\underline{y})= 
  \int_{-\infty}^\infty \frac{d\w}{(2\pi)} \: 
  \int \frac{d^{d-1}k_a}{(2\pi)^{d-1}}\:
  \frac{1}{\w}\,\tpsistar_{\uk}(\w,\ux)\,\tpsi_{\uk}(\w,\uy).
\end{equation}
and is therefore
\begin{eqnarray}
  \langle 0 | \, \phi(\t,z,x^a) \, \phi(\t,z',x^a) \, | 0
  \rangle&=& \frac{1}{\pi}\int_{-\infty}^\infty \frac{d\w}{(2\pi)}
  \int \frac{d^{d-1}k_a}{(2\pi)^{d-1}}
  \Bigl[e^{\frac{\pi \w}{2}} \Theta_x(R) +  e^{-{\frac{\pi \w}{2}}} \Theta_x(L)\Bigr] 
  \nonumber \\
  & &  \Bigl[e^{\frac{\pi \w}{2}} \Theta_y(R) +  e^{-{\frac{\pi \w}{2}}} \Theta_y(L)\Bigr]
  K_{i\w}(qz)\,K_{i\w}(qz')
\end{eqnarray}
where again we have only considered points separated  in the $x^1$ direction.
We can evaluate these integrals using the results of appendix C to give
\begin{equation}
  \label{eq:vevphiphiLRmer}
     \langle 0 | \, \phi(x^1) \, \phi(y^1) \, | 0 \rangle =
     \frac{1}{2\pi}\:
     \left(\frac{m}{2\pi\Delta x}\right)^{\frac{d-1}{2}}
     \,K_{\frac{d-1}{2}}(m\Delta x)
\end{equation}
where 
\begin{equation}
  \Delta x = \left\{\begin{array}{llcl}
      |z-u|&x,y \in R, R&\mbox{or}&x,y \in L, L \\
      |z+u|&x, y\in R, L&\mbox{or}& x, y\in L, R
      \end{array}\right.
\end{equation}

\begin{figure}[htb]
\centerline{\epsfxsize=3in \epsfbox{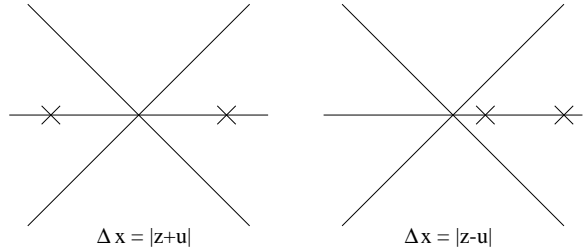}}
\caption{Distance between two points on the $t=\tau=0$ spacelike hypersurface.}
\label{fig:dx}
\end{figure}
As can be seen, Fig (\ref{fig:dx}) this is equivalent 
to the Minkowski two point function. This is true on any 
$\tau={\rm const}$ hypersurface because $\Delta x$ is 
just the geodesic distance between the two points and
exactly equals the  geodesic distance between the same 
two points in Minkowski spacetime.

A similar calculation (see appendix B) gives the VEV of
$T_{\tau \tau}$ with point--split arguments,
\begin{equation}
  \langle 0 | T_{\tau \tau}(z,z') | 0 \rangle =
  - z^2 \left(\frac{m}{2\pi |z-z'|}\right)^{\frac{d+1}{2}}\:
K_{\frac{d+1}{2}}(m|z-z'|)
\end{equation}
where the two points are in the same wedge. This again 
is identical to the Minkowski spacetime result, up to
coordinate transformation factors.

In particular, notice that these results are quite different from those
found for the single Rindler wedge {\bf R}. A correct foliation of Minkowski
spacetime must be based on spacelike hypersurfaces which are complete
Cauchy surfaces for the whole spacetime.

\vskip0.3cm
We see, therefore, that despite the radically different choice of foliations,
viz.~equal-$t$ surfaces or Rindler-Milne,
both the Wightman functions and energy-momentum tensor expectation values
are identical. 
This provides impressive evidence that, in general, physical quantities 
will indeed be foliation  independent, even though the vacuum wave functionals
themselves necessarily depend on the foliation chosen to implement the
{\schr} picture. 

\section*{Acknowledgements}

One of us (GMS) would like to thank Prof.~J-M.~Leinaas for extensive
discussions and hospitality at the University of Oslo and the Norwegian
Academy of Sciences. We are both grateful to Dr.~W.~Perkins 
for many helpful discussions. 

\newpage
\appendix
%%%%%%%%%%%%%%%%%%%%%%%%%%%%
\section{Energy-momentum tensor in the Milne universe}

In this appendix we calculate the expectation value of
the `energy' component of the energy-momentum tensor in the
comoving and conformal vacua. The expectation value is the coincidence
limit of (\ref{eq:energytensor}), which in terms of the Fourier transformed
kernel (\ref{eq:milnekernel2}) is
\[
  \langle0| T_{zz}(\t,\t';z)|0\rangle = \frac{1}{2} \!
  \int_{-\infty}^\infty \!\!\frac{d\w}{2\pi} \!\!
  \int \!\!\frac{d^{d-1}k_a}{(2\pi)^{d-1}}
  e^{i\w(\t-\t')}\frac{\left\{
      \widetilde{G}(\uk;z)\widetilde{G}^\ast(-\uk;z) + \frac{1}{z^2}
      \left[\frac{\w^2}{z^2}+q^2\right]\right\}}{\left\{
      \widetilde{G}(\uk;z) + \widetilde{G}^\ast(-\uk;z)    \right\}}
\]
Here, we are working in $d$ space dimensions, as needed in section 5,
and have point-split in the $x^1$ direction only.

\vskip0.2cm

The kernel for the comoving vacuum is specified by choosing the wave equation
solution $\tilde\psi(z,\w) = H^2_{i\w}(qz)$. The expectation value in this 
vacuum is therefore
\[
%  \langle0| T_{zz}|0\rangle_{\rm COM} =
\frac{\pi}{8}  
\int_{-\infty}^{\infty} \frac{d\w}{2\pi}
\int \frac{d^{d-1}k_a}{(2\pi)^{d-1}}
  e^{i\w(\t-\t')}\biggl\{\Bigl[\partial_z H^1_{i\w}(qz)\Bigr]
\Bigl[\partial_z H^2_{i\w}(qz)\Bigr]
    +\biggl[\frac{\w^2}{z^2}+q^2\biggr]H^1_{i\w}(qz)H^2_{i\w}(qz)
  \biggr\}
\]
where we have used the Wronskian $W[\tpsi,\tpsi^\ast] =\tpsi[\partial_z\tpsi^\ast] 
-[\partial_z \tpsi]\tpsi^\ast = -\frac{4i}{\pi z}$. Using standard
properties of derivatives of Hankel functions, this can be rewritten as
\begin{eqnarray}
&&\hspace*{-1.3cm}\frac{\pi}{8}  
\int_{-\infty}^{\infty} \frac{d\w}{2\pi}
\int \frac{d^{d-1}k_a}{(2\pi)^{d-1}}
  e^{i\w(\t-\t')}\biggl\{q^2\Bigl[H^1_{i\w}(qz)H^2_{i\w}(qz)-
H^1_{i\w+1}(qz)H^2_{i\w-1}(qz)\Bigr] +\nonumber \\
&&\hspace*{-0.7cm}\frac{2\w^2}{z^2}H^1_{i\w}(qz)H^2_{i\w}(qz) 
+\frac{i\w q}{z}\Bigl[H^1_{i\w}(qz)H^2_{i\w-1}(qz)+
H^1_{i\w+1}(qz)H^2_{i\w}(qz)\Bigr]
  \biggr\}
\label{eq:toocomcalc}
\end{eqnarray}
Expressing the Hankel functions as modified Bessel functions and performing
the $\w$ integration (using eq.(\ref{eq:besselintegral2}) and derivatives of it with respect to $\rho$)
reduces this to 
\[
 -\frac{1}{2\pi \sigma} \int 
 \frac{d^{d-1}k_a}{(2\pi)^{d-1}} \sqrt{k_a^2+m^2}\:K_1(\sqrt{k_a^2+m^2}\,\sigma)
\]
Finally, therefore, we find
\begin{equation}
  \label{eq:toocom}
  \langle0| T_{zz}(\t,\t';z)|0\rangle_{\rm COM} =
  - \left(\frac{m}{2\pi \sigma}\right)^{\frac{d+1}{2}}\:
  K_{\frac{d+1}{2}}(m\sigma)
\end{equation}
with $\sigma$ is the geodesic interval between the two point-split points.
The coincidence limit ($\tau \rightarrow \tau'$)  is of course  divergent.
 
An alternative representation of the expectation value is found by taking the 
coincidence limit before performing the $\w$ integration. Working
in $(1+1)$ dimensions as in section 2 gives 
\[
  \langle0| T_{zz}(z)|0\rangle_{\rm COM} = \int_0^\infty \frac{d\w}{2\pi}
\bigg\{\frac{\w}{z^2} + \frac{\pi m^2}{4}\Bigl[
H^1_{i\w}(mz)H^2_{i\w}(mz)  - H^1_{i\w+1}(mz)H^2_{i\w-1}(mz) 
\Bigr]\bigg\}
\]
Reexpressing the Hankel functions in terms of  Bessel functions 
it can be shown that 
\begin{eqnarray}
  \label{eq:toocom2}
  &&\hspace*{-1cm}\langle0| T_{zz}(z)|0\rangle_{\rm COM} = 
  \int_0^\infty \frac{d\w}{2\pi}
  \frac{\w \cosh(\pi \w)}{z^2 \sinh(\pi \w)}
  +\int_0^\infty \frac{d\w}{2\pi} \frac{\pi m^2}{4 \sinh^2(\pi \w)}\times 
  \nonumber \\
  &&\bigg\{ 2 \cosh(\pi \w)\Bigl[J_{i\w}(mz)J_{-i\w}(mz)+
  J_{i\w+1}(mz)J_{-i\w+1}(mz)\Bigr]-[J_{-i\w}(mz)]^2\nonumber \\
  && -[J_{i\w}(mz)]^2+J_{i\w-1}(mz)J_{i\w+1}(mz)
  +J_{-i\w-1}(mz)J_{-i\w+1}(mz)\bigg\}
\end{eqnarray}
In the limit of small $z$, only the first term is of  
$O(z^{-2})$, the others being of $O(z^{0})$.
The first term is also $m$ independent, while the others are of $O(m^2)$.

\vskip0.3cm
The corresponding calculation of the expectation value of the
$T_{zz}$ component of the energy-momentum tensor in the conformal
vacuum, defined by specifying $\tpsi(z,\w) = J_{-i|\w|}(mz)$, 
gives
\[
\frac{\pi}{2} \int_0^\infty \frac{d\w}{2\pi} \frac{1}{\sinh(\pi \w)}
\bigg\{\Bigl[\partial_z J_{i\w}(mz)\Bigr]\Bigl[\partial_z J_{-i\w}(mz)\Bigr]
+\Bigl[\frac{\w^2}{z^2} + m^2\Bigr] J_{i\w}(mz) J_{-i\w}(mz)\biggr\}
\]
where $W[\tpsi,\tpsi^\ast] =\frac{2i\sinh(\pi \w)}{\pi z}$. 
In this case, we find
\begin{eqnarray}
  \label{eq:tooconf}
  &&\hspace*{-1cm}\langle0| T_{zz}(z)|0\rangle_{\rm CONF}= 
  \int_0^\infty   \frac{d\w \:\w}{2\pi z^2} \nonumber \\
  && +\int_0^\infty  \frac{d\w \: m^2}{4\sinh(\pi \w)}\biggl\{
   J_{i\w}(mz) J_{-i\w}(mz)+  J_{i\w+1}(mz) J_{-i\w+1}(mz)
   \biggr\}
\end{eqnarray}
Again, the first term dominates in the small $z$ or small mass limits, 
being the only one of $O(z^{-2})$ or independent of $m$.

%%%%%%%%%%%%%%%%%%%%%%%%%%%%
\section{Energy-momentum tensor in Rindler spacetime}

In this appendix we calculate the VEV of the 
energy component of the energy--momentum tensor in the {\bf R}
Rindler wedge and in the {\bf L} and {\bf R} wedges together. The
expectation value is the coincidence limit of (\ref{eq:energytensor}),
\[
\langle0| T_{\tau \tau}(\ux,\uy)|0\rangle = \frac{z^2}{4}
\Bigl\{ G(\ux,\uy) + \Big[\frac{\partial^2}{\partial z \partial z'}
+\frac{\partial^2}{\partial x^a \partial y^a} +m^2 
\Big] \Delta(\ux,\uy)\Bigr\}
\]
Again we are working in $d$ space dimensions and shall point--split
in the $x^1$ direction only.

\vskip0.3cm

In the {\bf R} Rindler wedge this expectation value reduces to
\[
\frac{z^2}{2\pi^2} \int_0^\infty d\w \int \frac{d^{d-1}k_a}{(2\pi)^{d-1}}
\left\{
\frac{\omega^2}{z z'} + \frac{\partial^2}{\partial z \partial z'}
+q^2\right\}  \sinh (\pi \omega) K_{i\omega}(qz) K_{i\omega}(qz')
\]
with $q^2= k_a^2 +m^2$ and where we have introduced the complete
orthonormal set of solutions to the Fourier transformed wave equation,
viz. (\ref{eq:wavesolutionLR}). Using standard properties of 
modified Bessel functions and performing the $\omega$ integration
gives
\begin{eqnarray}
  &&\hspace*{-1cm}\frac{z^2}{2\pi}\int 
  \frac{d^{d-1}k_a}{(2\pi)^{d-1}}\biggl\{
  -\frac{q K_1(q|z-z'|)}{|z-z'|} + 
  \int_0^\infty \frac{dv}{(v^2+\pi^2)} \biggl[\frac{P(v)K_0(qZ)}{z z'}
  \nonumber \\
  &&\hspace*{4.5cm} + \frac{q}{Z} (2+\cosh v)K_1(qZ) - Q(v) q^2 K_2(qZ)
  \biggr]\biggr\}\nonumber 
\end{eqnarray}
where 
\begin{eqnarray}
  P(v)&=&{2(3v^2-\pi^2)}{(v^2+\pi^2)^{-2}}\nonumber \\
  Q(v) &=& 1 + (z+z'\cosh v)(z'+z\cosh v)Z^{-2}\nonumber 
\end{eqnarray}
Finally we find the expectation value in the {\bf R} Rindler wedge is
\begin{eqnarray}
  &&\langle 0|T_{\t\t}(z,z')|0\rangle
  = -z^2\biggl(\frac{m}{2\pi |z-z'|}\biggr)^{\frac{d+1}{2}} 
  K_{\frac{d+1}{2}}(m|z-z'|)
  +\int_0^\infty \frac{dv \:z^2}{2\pi (v^2+\pi^2)}\times \nonumber \\
  &&\hspace*{4cm}\mbox{$\biggl[ 2\pi\Bigl(2 + \cosh v -(d+1) Q(v)\Bigr)
    \biggl(\frac{m}{2\pi Z}\biggr)^{\frac{d+1}{2}}K_{\frac{d+1}{2}}(mZ)$}
  \nonumber \\
  &&\hspace*{5cm}\mbox{$   +
    \Bigl(\frac{P(v)}{z z'}-Q(v)m^2\Bigr)
    \biggl(\frac{m}{2\pi Z}\biggr)^{\frac{d-1}{2}}
    K_{\frac{d-1}{2}}(mZ)\biggr]$}
\end{eqnarray}

\vskip0.3cm

In considering the expectation value in the {\bf L} and {\bf R} Rindler
wedges together (as in section 5) we use the complete
orthonormal set of solutions to the Fourier transformed wave equation,
(\ref{eq:waveLandR}) which gives
\begin{eqnarray}
 && \hspace*{-1cm}\frac{z^2}{4\pi^2} \int_{-\infty}^\infty d\w 
  \int \frac{d^{d-1}k_a}{(2\pi)^{d-1}}
  \left\{\frac{\omega^2}{z z'} + 
    \Bigl[\frac{\partial^2}{\partial z \partial z'}+q^2\Bigr]\right\}  
   K_{i\omega}(qz) K_{i\omega}(qz') \times \nonumber \\
   &&[e^{\pi \omega} \Theta_x(R)\Theta_y(R)+ \Theta_x(R)\Theta_y(L) 
   +\Theta_x(L)\Theta_y(R)+e^{-\pi \omega} \Theta_x(L)\Theta_y(L)]
   \nonumber
\end{eqnarray}
As we require the coincidence limit we shall only consider
points separated in the same wedge. Implementing this and 
performing the $\omega$ integration gives
\[
- \frac{z^2}{2\pi} \int \frac{d^{d-1}k_a}{(2\pi)^{d-1}}
\frac{q}{|z-z'|} K_1(q|z-z'|)
\]
which results in 
\begin{equation}
  \langle 0 | T_{\tau \tau}(z,z') | 0 \rangle =
  - z^2 \left(\frac{m}{2\pi |z-z'|}\right)^{\frac{d+1}{2}}\:
  K_{\frac{d+1}{2}}(m|z-z'|)
\end{equation}

%%%%%%%%%%%%%%%%%%%
\section{Integrals}

The following integrals arise in the calculations of expectation values:
\begin{eqnarray}
  \label{eq:intI}
  \int_{0}^\infty dx\: K_{ix}(a) K_{ix}(b)&=& \frac{\pi}{2} K_0(a+b) \\
  \label{eq:intII}
  \int_{0}^\infty dx\:x^2 K_{ix}(a) K_{ix}(b)&=& \frac{\pi}{2} 
  \frac{a b}{(a+b)} K_1(a+b) \\
  \label{eq:intIII}
  \int_{0}^\infty dx\: \cosh({\pi x}) K_{ix}(a) K_{ix}(b) 
  &=& \frac{\pi}{2} K_0(|a-b|)\\
  \label{eq:intIV}
  \int_{0}^\infty dx\:x^2 \cosh({\pi x}) K_{ix}(a) K_{ix}(b) 
  &=&- \frac{\pi}{2} \frac{a b}{|a-b|}  K_1(|a-b|)\\
  \label{eq:intV}
  \int_0^\infty dx \:\sinh (\pi x) K_{ix}(a) K_{ix}(b) &=& \frac{\pi}{2}
  K_0(|a-b|)  \\
  &&\hspace*{-1.5cm} -\pi \int_0^\infty \frac{dz}{\pi^2+z^2}\: 
  K_0(\mbox{$\sqrt{a^2\!+\!b^2\!+\! 2ab\cosh z}$})\nonumber\\
  \label{eq:intVI}
  \int_0^\infty dx \:x^2 \sinh (\pi x) K_{ix}(a) K_{ix}(b) &=& 
  - \frac{\pi}{2}  \frac{a b}{|a-b|}   K_1(|a-b|)  \\
  &&\hspace*{-1.5cm} +
  2 \pi \int_0^\infty \frac{dz\:(3z^2-\pi^2)}{(\pi^2+z^2)^3}\: 
  K_0(\mbox{$\sqrt{a^2\!+\!b^2\!+\!2 a b \cosh z}$})\nonumber
\end{eqnarray}
valid for $a, b>0$ \cite{grad80, hill86}. 

The Mellin transform \cite{magnus66} of the Bessel function,
with $\mbox{Re}\,(s,\alpha,\beta)>0$ is
\begin{equation}
  \label{eq:mellintransform}
  \int_0^\infty dx\,x^{s-1}\,(x^2+\beta^2)^{-\frac{\nu}{2}}\: 
  K_\nu(\alpha\sqrt{x^2+\beta^2}) = a^{-\frac{s}{2}}\,2^{\frac{s}{2}-1}\,
  \beta^{\frac{s}{2}-\nu}\,\Gamma(\mbox{$\frac{s}{2}$})\:
  K_{\frac{s}{2}-\nu}(\alpha \beta).
\end{equation}
We also need \cite{grad80}, 
\begin{eqnarray}
  \label{eq:besselintegral}
  \int_{-\infty}^{\infty} \!\!\!dx e^{i\rho x}
  K_{ix+\nu}(a)K_{\nu-ix}(b)\!\! &=&\!\!
  \pi \mbox{$
    \left(\frac{a e^\rho + b}{a + b e^\rho}\right)^\nu$}
K_{2\nu}(\mbox{$\sqrt{a^2\!\! +\! b^2\!\! +\! 2ab \cosh \rho}$})\\
  \label{eq:besselintegral2}
  \int_{-\infty}^{\infty} \!\!\!dx e^{i\rho x}
  K_{ix+\nu}(ia)K_{ix+\omega}(-ia)\!\!&=&\!\!
  \pi e^{-\frac{\rho}{2}(\omega+\nu)} e^{i\frac{\pi}{2}(\nu-\omega)}
  K_{\nu-\omega}(2\alpha \sinh\mbox{$\frac{\rho}{2}$})
\end{eqnarray}
where $|\mbox{arg} a|+|\mbox{arg} 
b|+|\mbox{Im} \rho|<\pi$.

\newpage
%%%%%%%%%%%%%%%%%%%%%%%%%%%%
%                          %
%      BIBLIOGRAPHY        %
%                          %
%%%%%%%%%%%%%%%%%%%%%%%%%%%%


\begin{thebibliography}{10}

\bibitem{long96}
D.~V. Long and G.~M. Shore, {preprint hep-th/9605004}, SWAT 96/56.

\bibitem{fulling73}
S.~A. Fulling, Phys. Rev. D7 (1973) 2850.

\bibitem{candelas76}
P.~Candelas and D.~J. Raine, J. Math. Phys. 17 (1976) 2101.

\bibitem{lee86}
T.~D. Lee, Nucl. Phys. B264 (1986) 437.

\bibitem{mcavity93}
D.~M. Mc{A}vity and H.~Osborn, Nucl. Phys. B394 (1993) 728.

\bibitem{hamazaki96}
T.~Hamazaki, M.~Sasaki, T.~Tanaka and K.~Yamamoto,
Phys. Rev. D53 (1996) 2045.

\bibitem{boulware75}
D.~G. Boulware, Phys. Rev. D11 (1975) 1404.

\bibitem{disessa74}
A.~{diSessa}, J. Math. Phys. 15 (1974) 1892.

\bibitem{sommerfield74}
C.~M. Sommerfield, Ann. Phys. 84 (1974) 285.

\bibitem{gromes74}
D.~Gromes, H.~J. Rothe and B.~Stech, Nucl. Phys. B75 (1974) 313.

\bibitem{magnus66}
W.~Magnus, F.~Oberhettinger and R.~P. Soni,
 {Formulas and Theorems for Special Functions of Mathematical
  Physics} (Springer-Verlag, Berlin, 1966).

\bibitem{grad80}
I.~S. Gradshteyn and I.~M. Ryzhik,
 {Table of Integrals, Series, and Products}
(Academic Press, New York, 1980).

\bibitem{bunch78}
T.~S. Bunch, S.~M. Christensen and S.~A. Fulling,
Phys. Rev. D18 (1978) 4435.

\bibitem{rindler66}
W.~Rindler, Am. J. Phys. 34 (1966) 1174.

\bibitem{unruh76}
W.~G. Unruh, Phys. Rev. D14 (1976) 870.

\bibitem{sciama81}
D.~W. Sciama, P.~Candelas and D.~Deutsch, Adv. Phys. 30 (1981) 327.

\bibitem{freese85}
K.~Freese, C.~T. Hill and M.~Mueller, Nucl. Phys. B255 (1985) 693.

\bibitem{hill85}
C.~T. Hill, Phys. Lett. B155 (1985) 343.

\bibitem{hill86}
C.~T. Hill, Nucl. Phys. B277 (1986) 547.

\bibitem{bell83}
J.~S. Bell and J.~M. Leinaas, Nucl. Phys. B212 (1983) 131.

\end{thebibliography}
\end{document}